\documentclass[
    ,final            
  ]
  {aipproc}

\layoutstyle{6x9}

\usepackage[]{amssymb}
\usepackage[]{rotating}

\begin{document}

\title{Radial Dependence of Extinction in Parent Galaxies of Supernovae}


\author{Bojan Arbutina}{
  address={Astronomical Observatory, Volgina 7, 11160 Belgrade, Serbia}
  ,altaddress={Department of Astronomy, Faculty of Mathematics, University of Belgrade, Studentski trg 16,  11000 Belgrade, Serbia}
}

\begin{abstract}
 The problem of extinction is the most important issue to be dealt with
 in the process of obtaining true absolute magnitudes of core-collapse supernovae (SNe).
 The plane-parallel model which gives absorption dependent on galaxy inclination, widely used in the past,
 was shown not to describe extinction adequately.
 We try to apply an alternative model which introduces radial dependence of extinction.
 A certain trend of dimmer SNe with decreasing radius from the center of a galaxy was found,
 for a chosen sample of stripped-envelope SNe.
\end{abstract}

\maketitle


\section{INTRODUCTION}

 As well known, the supernovae (SNe) Ia are widely used by astronomers as distance indicators
 because of their small dispersion in peak absolute magnitude. SNe II, on the other hand, are a quite heterogenous
 class with absolute magnitudes $-19 \lesssim \mathrm{M_B} \lesssim
 -14$ (see e.g. \cite{1}). The situation is still unclear regarding stripped-envelope
 SNe (Ib/c). The progenitors of these SNe are massive stars that have lost most or
all of their hydrogen envelopes, by strong winds such as in
Wolf-Rayet stars or through mass transfer to a companion star in
Roche lobe overflow or a common envelope phase. We have focused on
finding the absolute B magnitude for these SNe (for a discussion
on V magnitudes see \citep{2}).

\vspace{-6mm}

\section{ANALYSIS AND RESULTS}

For the absolute magnitude at maximum (blue) light we can
generally write
\begin{equation}
\mathrm{M_B^0} = \mathrm{m_B} - \mu - \mathrm{A}_G - \mathrm{A}_g
= \mathrm{M_B} - \mathrm{A}_g
\end{equation}
where $\mathrm{m_B}$ is apparent magnitude, $\mathrm{M_B}$ is
absolute magnitude uncorrected for extinction in parent galaxy,
$\mu = 5\log d\mathrm{[Mpc]} +25$ is distance modulus,  $d$ is the
distance to the galaxy, $\mathrm{A}_G$ and $\mathrm{A}_g$ are
Galactic extinction and extinction in parent galaxy, respectively.
The problem of extinction is the most important issue to be dealt
with in the process of obtaining true absolute magnitudes of
core-collapse (including stripped-envelope) SNe.
 The plane-parallel model which gives absorption dependent on galaxy inclination $i$, $\mathrm{A}_g=\mathrm{A}_o
\mathrm{sec}\ i$, widely used in the past,
 was shown not to describe extinction adequately \cite{3}.
 We try to apply an alternative model which introduces radial dependence of
 extinction \cite{4}.

 Bearing in mind the short life of their progenitors, we may
 assume that the stripped-envelope SNe, we are interested in, are practically in the galactic plane
 ($z=0$). The radial position of a supernova in a galaxy is then
\begin{eqnarray}
& r^2 = d^2\big( (x')^2+ (y')^2 \sec ^2 i \big) = d^2 \big(x^2+y^2\big) \cdot & \nonumber \\
& \cdot \big(\cos ^2 (\arctan ({y}/{x}) + \Pi -90^{\circ}) + \sin
^2 (\arctan ({y}/{x}) + \Pi -90^{\circ})\sec ^2 i \big), &
\end{eqnarray}
where $x$ and $y$ give SN offset from the center of the galaxy in
radians and $\Pi$ is the position angle of the major axis (see
Fig. 1). If not given, offsets ($x$, $y$) can be calculated from
rectascension ($\alpha$) and declination ($\delta$); $x \approx
(\alpha _{\mathrm{SN}} - \alpha _{g})\cos \delta _{g}$, $y \approx
(\delta _{\mathrm{SN}} - \delta _{g})$.


\begin{figure}
  \includegraphics[width=.85\textwidth]{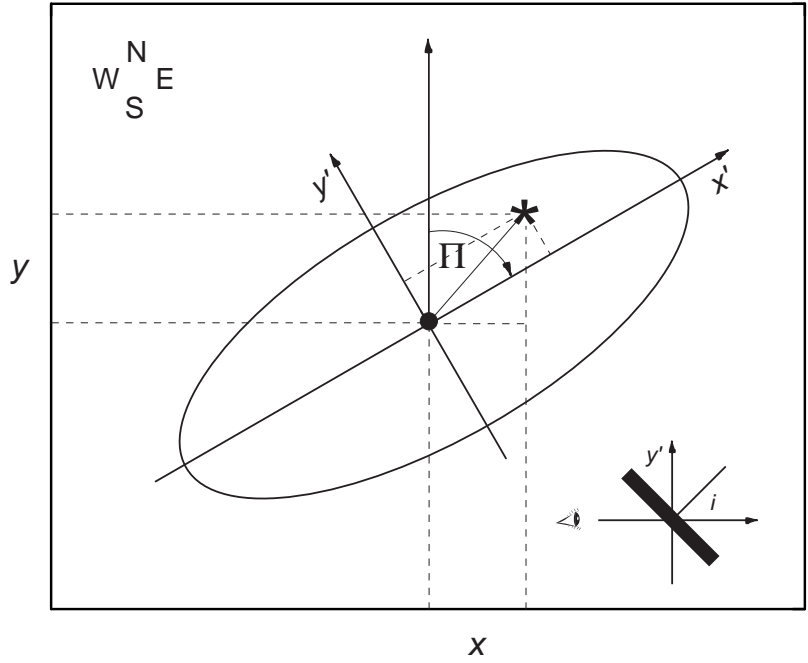}
   \caption{Calculating radial position of a supernova in its parent galaxy.}
\end{figure}

Table 1 gives a sample of SNe Ib/c with known peak B magnitude
from the Asiago Supernova Catalogue (ASC) \cite{5}. Fig. 2 shows
peak magnitude uncorrected for parent galaxy extinction against
the radial position of SN $r$ in the units of galactic radius
$R=D/2$. We see that there is a certain trend of dimmer SNe with
decreasing radius \cite{6}. If we assume that
$\mathrm{A}_g=\mathrm{A}_o e ^{-\alpha_o r/R}$ then
\begin{equation}
\mathrm{M_B} = \mathrm{M_B^0} + \mathrm{A}_o e ^{-\alpha_o r/R}.
\end{equation}
Fit to data gives SNe Ib/c intrinsic absolute magnitude
 \begin{equation}
\mathrm{M_B^0} = -17.58 \pm 0.27.
\end{equation}

\begin{sidewaystable}
 \begin{minipage}{\textheight}
 {\small
  {{\bf TABLE 1.} SNe Ib/c with listed B magnitude at the time of maximum from the
  Asiago Supernova Catalogue (ASC). We have excluded SN 1954A in
  irregular galaxy, 1966J since it was shown to be
  SN Ia \cite{7}, and peculiar SNe Ic (Id) 1998bw and 2002ap.
  All data are from ASC, except for distance moduli (Nearby
  Galaxy Catalogue - NGC \cite{8}) and correction for Galactic absorption $\mathrm{A}_G$ which is from
  RC3 catalogue \cite{9}.
  Parent galaxy extinction is omitted since we found significant
  discrepancy for $\mathrm{A}_o$ in RC3 and NGC. $\mathrm{M_B}$ is
  absolute magnitude uncorrected for extinction in the parent galaxy.
  }
  \vskip 5mm
  \centering
  \begin{tabular}{@{\extracolsep{-0.2mm}}lclcrrrrrrrrrrrrrrrrrrrrr@{}}
  \hline
   Supernova   & SN & Galaxy & Galaxy &
   \multicolumn{3}{c}{Distance}&
   \multicolumn{3}{c}{Inclination}&
   \multicolumn{3}{c}{Diameter}&
   \multicolumn{3}{c}{SN radial}&
   \multicolumn{3}{c}{Apparent}&
   \multicolumn{3}{c}{Galactic}&
   \multicolumn{3}{c}{Absolute}\\
           &  type &  & type &
   \multicolumn{3}{c}{modulus}&
   \multicolumn{3}{c}{}&
   \multicolumn{3}{c}{}&
   \multicolumn{3}{c}{position}&
   \multicolumn{3}{c}{magnitude}&
   \multicolumn{3}{c}{absorption}&
   \multicolumn{3}{c}{magnitude}\\
              & & & &
   \multicolumn{3}{c}{$\mu$}&
   \multicolumn{3}{c}{$i\ \mathrm{[^{\circ}]}$}&
   \multicolumn{3}{c}{$D\ \mathrm{[kpc]}$}&
    \multicolumn{3}{c}{$r\ \mathrm{[kpc]}$}&
   \multicolumn{3}{c}{$\mathrm{m_B}$}&
   \multicolumn{3}{c}{$\mathrm{A}_G$}&
   \multicolumn{3}{c}{$\mathrm{M_B}$}\\
\hline
SN 1972R & Ib & NGC 2841 & Sb && 30.39 &&\hspace{2mm} & 65 &&& 27 &&& 10.7 &&&  12.85  &&\hspace{1mm} &  0    &&& -17.54 &\\
SN 1983N & Ib & NGC 5236 & SBc&& 28.35 &&\hspace{2mm} & 21 &&& 18 &&& 3.8  &&&  11.70  &&\hspace{1mm} &  0.15 &&& -16.80 &\\
SN 1984I & Ib &ESO 393-99&SBcd&& 33.48 &&\hspace{2mm} & 25 &&& 30 &&& 11.0 &&&  16.60  &&\hspace{1mm} &  0.45 &&& -17.33 &\\
SN 1984L & Ib & NGC 991  & SBc&& 31.37 &&\hspace{2mm} & 28 &&& 16 &&& 3.5  &&&  14.00  &&\hspace{1mm} &  0    &&& -17.37 &\\
SN 2000H & Ib & IC 454   &SBab&& 33.89 &&\hspace{2mm} & 58 &&& 30 &&& 9.1  &&&  17.90  &&\hspace{1mm} &  1.44 &&& -17.43 &\\
SN 1962L & Ic & NGC 1073 & SBc&& 30.91 &&\hspace{2mm} & 25 &&& 21 &&& 5.7  &&&  13.94  &&\hspace{1mm} &  0.07 &&& -17.04 &\\
SN 1983I & Ic & NGC 4051 &SBbc&& 31.15 &&\hspace{2mm} & 35 &&& 26 &&& 5.5  &&&  13.70  &&\hspace{1mm} &  0    &&& -17.45 &\\
SN 1983V & Ic & NGC 1365 & SBb&& 31.14 &&\hspace{2mm} & 58 &&& 54 &&& 6.8  &&&  14.67  &&\hspace{1mm} &  0    &&& -16.47 &\\
SN 1987M & Ic & NGC 2715 & SBc&& 31.55 &&\hspace{2mm} & 74 &&& 28 &&& 2.1  &&&  15.30  &&\hspace{1mm} &  0.02 &&& -16.27 &\\
SN 1991N & Ic & NGC 3310 &SBbc&& 31.36 &&\hspace{2mm} & 19 &&& 15 &&& 0.8  &&&  15.50  &&\hspace{1mm} &  0    &&& -15.86 &\\
SN 1994I$^\dag$ & Ic & NGC 5194 & Sbc && 29.62 &&\hspace{2mm} & 48 &&& 24 &&& 1.1  &&&  13.77  &&\hspace{1mm} &  0 &&& -15.85 &\\
\hline

\end{tabular}
} \begin{flushleft}
  {\footnotesize $^\dag$ Peak B magnitude and distance modulus
  were adopted from Ref. \cite{10}.}
  \end{flushleft}
\end{minipage}
\end{sidewaystable}

\begin{figure}
  \includegraphics[width=.85\textwidth]{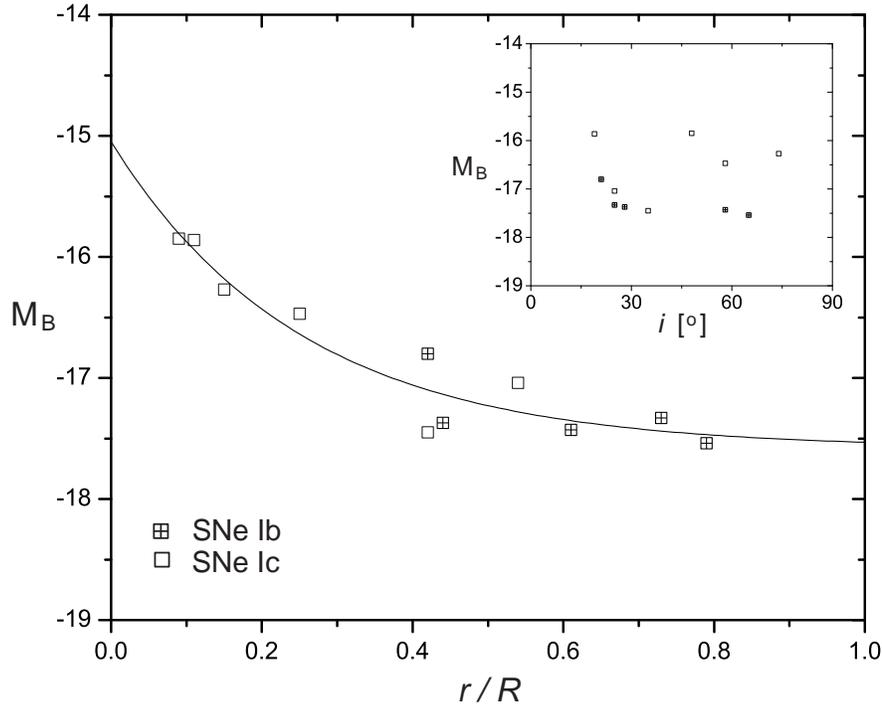}
   \caption{SN absolute magnitude, uncorrected for the parent galaxy extinction, is plotted against relative radial position of a SN in the galaxy. The curve represents the best fit (see the text).
  The inset in the upper-right corner of the plot shows, again, magnitude only now as the function of the galaxy inclination. There is no apparent dependence of $\mathrm{M_B}$ on $i$.
  }
\end{figure}


\begin{theacknowledgments}
  The work on this paper was financially supported by the
Ministry of Science of Serbia through the projects No. 146003 and
146012.
\end{theacknowledgments}



\bibliographystyle{aipproc}   


\vspace{-1mm}

\begin{thebibliography}{9}

\bibitem{1} D. L. Miller and D. Branch, {\it AJ}, {\bf 100}, 530 (1990).

\bibitem{2} D. Richardson, D. Branch, and E. Baron, {\it AJ}, {\bf 131}, 2233 (2006).

\bibitem{3} E. Cappellaro, M. Turatto, D. Yu. Tsvetkov, O. S. Bartunov, C. Pollas, R. Evans and M. Hamuy, {\it A\&A}, {\bf 322},
431 (1997).

\bibitem{4} K. Hatano, D. Branch and J. Deaton, {\it ApJ}, {\bf
502}, 177 (1998).

\bibitem{5} R. Barbon, V. Boundi, E. Cappellaro and M. Turatto, {\it
A\&A}, {\bf 139}, 531 (1999).

\bibitem{6} B. Arbutina, M.Sc. Thesis, University of Belgrade
(2005).

\bibitem{7} G. Casebeer, D. Branch, M. Blaylock, J. Millard, E. Baron,
D. Richardson and C. Ancheta, {\it PASP}, {\bf 112}, 1433 (2000).

\bibitem{8} R. B. Tully, {\it Nearby Galaxy Catalogue}, Cambridge
University Press, Cambridge (1988).

\bibitem{9} G. de Vaucouleurs, A. de Vaucouleurs, H. G. Corwin, R. J.
Buta, G. Paturel and P. Foque, {\it Third Reference Catalogue of
Bright Galaxies}, Springer-Verlag, New York (1991).

\bibitem{10} M. W. Richmond et al., {\it AJ}, {\bf 111}, 327 (1996).


\end{thebibliography}





\end{document}